\documentclass[nofootinbib,notitlepage,twocolumn,superscriptaddress,10pt,aps,prd]{revtex4-1}

\usepackage{graphicx,epsfig,amssymb,times} 
\usepackage{amsmath,amsfonts}
\usepackage{bm}
\usepackage{epstopdf}
\usepackage[caption=false]{subfig}
\usepackage[usenames]{color}     
\usepackage{tikz}
\usetikzlibrary{decorations.pathmorphing}
\usepackage{float}

\definecolor{coolblack}{rgb}{0.0, 0.18, 0.39}
\definecolor{darkred}{rgb}{0.5,0,0}
\definecolor{darkgreen}{rgb}{0,0.5,0}
\definecolor{darkblue}{rgb}{0,0,0.5}
\definecolor{lapislazuli}{rgb}{0.15, 0.38, 0.61}
\definecolor{venetianred}{rgb}{0.78, 0.03, 0.08}
\definecolor{bleudefrance}{rgb}{0.19, 0.55, 0.91}
\definecolor{dogwoodrose}{rgb}{0.84, 0.09, 0.41}

\newcommand{\nt}[1]{\textcolor{magenta}{#1}}
\newcommand{\stout}[1]

\def\be{\begin{equation}}
\def\ee{\end{equation}}
\def\p{\partial}
\newcommand{\bea}{\begin{eqnarray}}
\newcommand{\eea}{\end{eqnarray}}
\def\l{\left}
\def\r{\right}
\newcommand{\lr}[1]{\left(#1\right)}

\newcommand{\tia}[1]{}

\newcounter{listcounter}

\begin{document}\title{\large Absorption by black hole remnants in metric-affine gravity}

\author{Adria Delhom}
\email{adria.delhom@uv.es}
\affiliation{Departamento de F\'isica Te\'orica and IFIC, 
Centro Mixto Universidad de Valencia - CSIC. 
Universidad de Valencia, Burjassot-46100, 
Valencia, Spain}
\author{Caio F. B. Macedo}\email{caiomacedo@ufpa.br}
\affiliation{Campus de Salin\'opolis, Universidade Federal do Par\'a,
68721-000, Salin\'opolis, Par\'a, Brazil}
\author{Gonzalo J. Olmo}
\email{gonzalo.olmo@uv.es}
\affiliation{Departamento de F\'isica Te\'orica and IFIC, 
Centro Mixto Universidad de Valencia - CSIC. 
Universidad de Valencia, Burjassot-46100, 
Valencia, Spain}
\author{Lu\'is C. B. Crispino}
\email{crispino@ufpa.br}
\affiliation{Faculdade de F\'{\i}sica, Universidade 
Federal do Par\'a, 66075-110, Bel\'em, Par\'a, Brazil.}

\begin{abstract}
Using numerical methods, we investigate the absorption properties of a family of nonsingular solutions {which arise in different metric-affine theories, such as quadratic and Born-Infeld gravity.} These solutions continuously interpolate between Schwarzschild black holes and naked solitons with wormhole topology. The resulting spectrum is characterized by a series of quasibound states excitations, associated with the existence of a stable photonsphere. 
\end{abstract}


\date{\today}

\maketitle


\section{Introduction}\label{sec:intro}

In the last years there has been increasing interest in the study of compact objects which may figure as astrophysical alternatives to classical black holes (BHs) or exhibit unconventional features, such as hair or signs of new high-energy physics~\cite{Johnson-McDaniel:2018uvs,Cunha:2017wao, Herdeiro:2017phl,Cardoso:2016oxy,Herdeiro:2014goa,Liebling:2012fv}. This interest has grown in parallel with the development of gravitational wave detectors, which have provided convincing evidence that collisions between massive astrophysical-size compact objects is a fact~\cite{AbbotGW, AbbotPM, GBM:2017lvd, Abbott:2017dke, Abbott:2017gyy,Barack:2018yly}. However, the current capabilities of such observatories are yet insufficient to confirm or rule out the existence of the BH event horizon itself and we will have to wait for future developments in order to have a chance to settle this issue, as well as other related questions. Therefore, the possibility to test subtle details of the strong gravity regime is still beyond our observational capabilities, and we must do our best to scrutinize the spectrum of phenomenological possibilities from a theoretical perspective. 

Among the various open questions posed by BH investigations, understanding whether spacetime singularities~\cite{Earman:1995fv,Geroch:1968ut,Berger:2002st,Wald:1997wa} are real or an artifact of our mathematical theories is one of the most challenging problems, from both technical and philosophical perspectives. Though the BH event horizon is taken by some authors as a possibility to minimize this issue, adopting an {\it out of sight, out of mind} attitude, a lot of effort has been devoted to the construction of nonsingular alternatives for BH interiors.  In this sense, the physical nature of singularities 
has been attacked from different perspectives in the literature, including non-linear corrections on the matter fields~\cite{Bardeen,ABG98,Bronnikov01} (see also Refs.~\cite{Ansoldi:2008jw,DiazAlonso:2009ak,DiazAlonso:2010eh,DiazAlonso:2012mb} for a general analysis of this issue), as well as non-perturbative effects \cite{Barcelo:2017lnx}, fully dynamical models of BH formation and evaporation~\cite{Fabbri:2005mw,Hayward:2005gi,Zhang:2014bea,Liu:2014kra,Malafarina:2017csn},  quantum-gravitational pressure counter-effects preventing the formation of the singularity~\cite{Rovelli:2014cta,Spallucci:2017aod,Abedi:2015yga,Ashtekar:2018lag}, or via the replacement of the event horizon by a compact surface mimicking the Schwarzschild radius as seen from far away observers~\cite{Cardoso:2016oxy,Bueno:2017hyj}.

We are interested in exploring some properties of a family of nonsingular BH solutions which arise when higher-curvature corrections to Einstein's gravity are considered in the metric-affine (or Palatini) approach. Rather than being designed in terms of exotic matter sources to exhibit or prevent certain properties, the solutions discussed here come out naturally from the gravitational field equations coupled to standard matter sources, what makes them more appealing and supports their naturalness. These solutions were first found by exploring semiclassical gravity effects on Reissner-Nordstr\"om BHs~\cite{Olmo:2012nx,Olmo:2011np}, but relaxing the compatibility between metric and connection in order to obtain ghost-free, second-order equations. Interestingly, they are also solutions of the so-called Eddington-inspired Born-Infeld theory of gravity~\cite{Olmo:2013gqa}, which has been the subject of intense study in the last few years (see Ref.~\cite{BeltranJimenez:2017doy} for a comprehensive review). The most remarkable property of these new solutions is that they represent geodesically complete spacetimes with wormhole structure~\cite{Olmo:2016fuc,Olmo:2015bya,Olmo:2015dba}, where a spherical throat replaces the troublesome central singularity typically found in General Relativity (GR). The standard GR solutions are recovered from the new ones in the appropriate limit, what puts forward that they are natural extensions generated by the ultraviolet new dynamics of the gravity model. 

Among the various families of solutions of this electrovacuum theory, there is a subset which is completely regular, in the sense that curvature invariants are bounded everywhere, even at the wormhole throat~\cite{Olmo:2013gqa,Olmo:2012nx,Olmo:2011np}. These solutions smoothly interpolate between Schwarzschild-like BHs (when their mass is sufficiently high) and naked solitons (when their mass approaches the Planck scale), always having a wormhole of finite area at their center. For this reason, because they smoothly connect massive BH solutions with Minkowski spacetime, they can be regarded as natural candidates for BH remnants~\cite{Lobo:2013ufa,Lobo:2014fma}. Remnants have been considered in the past as a possible way out for the information loss problem in the process of BH evaporation~\cite{Chen:2014jwq}. The ones we are considering here are particularly interesting because: (i) they represent regular spacetimes with no incomplete geodesics (information is not destroyed anywhere), (ii) their wormhole structure allows to send information that falls during the BH phase into another universe (via a white hole), and (iii) correlations between a pair member that falls into the BH and a partner that remains outside can be restored once the event horizon disappears, what avoids the need to store all the quantum information in the remnant itself. 

Leaving aside the specific context in which these solutions arise, this unconventional family of massive topological entities offers an interesting environment to study qualitative new features of BH remnants. It is with this idea in mind that we initiate an in depth analysis of the interaction of this type of objects with scalar waves. Given that their BH phase is essentially identical to that corresponding to Schwarzschild BHs~\cite{Olmo:2012nx,Olmo:2011np}, here we focus on the horizonless configurations (naked solitonic phase), which can be seen as two copies of Minkowski spacetime connected by a spherical wormhole, where the energy density concentrates. Due to the fact that these objects are horizonless alternatives to standard BHs and that they usually present photonspheres, we shall consider them as ECOs (extreme/exotic compact objects)~\cite{Cardoso:2017cqb}.

ECOs can be further characterized into subclassses, namely UCOs (ultra-compact objects) and ClePhOs (clean photonsphere objects)~\cite{Cardoso:2017cqb}. UCOs are compact objects with a photonsphere and ClePhOs are UCOs with effective radius very close to the Schwarzschild radius. It was recently found in Ref.~\cite{Macedo:2018yoi} that, due to an ``effective cavity" between ClePhOs' effective surface and its photonsphere, ClePhOs present an absorption spectrum characterized by Breit-Wigner like resonances which could allow for experimental searches. It is natural to extend the analysis that was done in Ref.~\cite{Macedo:2018yoi} in order to study the absorption properties of other types of ECOs existent in alternative theories, in search of characteristic signatures that could distinguish them from regular BHs or those treated in Ref.~~\cite{Macedo:2018yoi}.

We obtain that the absorption spectrum of the family of regular solutions studied here exhibits a pattern associated to a rich structure of quasibound states in the remnant phase, which distinguishes ECOs from regular BHs of the same mass. This structure of quasibound states is  similar to that found for ClePhOs~\cite{Macedo:2018yoi}. 
	
The remaining of this paper is organized as follows. In Sec.~\ref{sec:framework} we present the framework and the solutions. The absorption properties are investigated in Sec.~\ref{sec:Abs}, where we also analyze the trapped modes. Our numerical results are exhibited in Sec.~\ref{sec:NR}. We end with our final remarks in Sec.~\ref{sec:discusscf}. Throughout this paper we use metric signature $(-,+,+,+)$ and natural units, such that $G=\hbar=c=1$.

\section{Framework and solutions}
\label{sec:framework}

The BH solutions we are going to study arise in the context of ultraviolet modifications of GR in a metric-affine scenario, in which metric and connection are regarded as independent geometrical entities. This approach has several advantages and peculiarities that make it radically different from the usual metric approach, in which the connection is forced to be compatible with the metric {\it a priori}. An independent variation of the affine connection has been advocated to render ghost-free theories of gravity even for higher order curvature Lagrangians \cite{BeltranJimenez:2017doy,Olmo:2011uz}. This has been recently proved to be the case for higher order curvature gravities which have a projective symmetry \cite{BeltranJimenez:2019acz}
 
 In Ricci-based gravity theories (RBGs), the symmetric part of the connection can always be formally written as the Christoffel symbols of an auxiliary metric $q_{\mu\nu}$, while its antisymmetric part is trivial if no explicit coupling between matter and connection is considered (minimally coupled bosonic fields) \cite{Afonso:2017bxr}. These theories can be represented by an action of the form 
\begin{equation}\label{RBGaction}
S(g_{ab},{\Gamma_{ab}}^c,\psi_m)=\int d^4 x\sqrt{-g} \mathcal{L}_{RBG}(g_{ab},R_{(ab)}) +S_m \ ,
\end{equation}
 where $R_{(ab)}$ denotes the symmetrized Ricci tensor\footnote{The fact that only the symmetrized Ricci tensor is included in the action is imposed by projective symmetry to ensure that they are ghost free \cite{BeltranJimenez:2019acz}.}, $g_{ab}$ the spacetime metric, and $\psi_m$ the matter fields that appear in $S_m$. Here $R_{ab}\equiv{R^c}_{acb}$, where the Riemann tensor is ${R^d}_{c ab}=\partial_a{\Gamma_{bc}}^d -\partial_b{\Gamma_{ac}}^d+{\Gamma_{ae}}^d{\Gamma_{bc}}^e-{\Gamma_{be}}^d{\Gamma_{ac}}^e$, and ${\Gamma_{ab}}^c$ is the connection. The relation between the spacetime metric $g_{ab}$ and the auxiliary metric $q_{ab}$ is determined by a model-dependent deformation matrix ${\Omega^a}_b$, which is a nonlinear function of the stress-energy tensor of the matter fields, via the relation $q_{ab}=g_{ac}{\Omega^c}_b$. The equations that govern the auxiliary metric can be written in a compact form in terms of $q_{ab}$ and become (see Ref.~\cite{Olmo:2013gqa} for details)
\begin{equation}
{G^a}_b(q)=\frac{\kappa^2}{|\Omega|^{\frac{1}{2}}}\left[{T^a}_b-{\delta^a}_b\left(\mathcal{L}_G+\frac{T}{2}\right)\right] \ ,
\end{equation} 
where ${G^a}_b(q)=q^{ac}G_{cb}(q)$ is the Einstein tensor of the metric $q_{ab}$ with one index raised. 
 Considering a spherically symmetric and static Maxwell electric field coupled to the quadratic action $\mathcal{L}_G=R+\alpha(-R^2/2+ R_{ab}R^{ab})$ or to the Born-Infeld like model $\sqrt{-g}\mathcal{L}_{BI}=\sqrt{|g_{ab}+\epsilon R_{ab}|}-\sqrt{-|g_{ab}|}$ (with $\alpha=\nt{-}\epsilon$), one finds the line element
\be\label{metric}
ds^2=-A(x) dt^2+\frac{1}{A(x) \mathcal{Z}_+^2(x)}dx^2+r^2(x)\left(d\theta^2+\sin^2\theta d\varphi^2\right),
\ee
where
\begin{align}
&A(x) \equiv \frac{1}{\mathcal{Z}_+(x)}\l[1-\frac{r_S}{r_c}\frac{(1+\delta_1 \, H(x))}{ z(x) \, \mathcal{Z}_-^{1/2}(x)}\r],\label{Adef}\\
&z(x) \equiv \frac{r(x)}{r_c}  \ , \  \mathcal{Z}_\pm(x) \equiv 1\pm\frac{1}{z^4(x)} \ , \ \label{mathcalZ}\\
&r^2(x)=\frac{1}{2}\left(x^2+\sqrt{x^4+4r_c^2}\right),\label{zofzeta}\\
&r_c \equiv \sqrt{l_\epsilon r_q} \ , \ 
\delta_1 \equiv \frac{1}{2r_S}\l[\frac{r_q^3}{l_\epsilon}\r]^{1/2} \ ,\label{defdelta1} 
r_q^2 \equiv 2 q^2 .
\end{align}
Here the $x$ coordinate, defined through Eq.~(\ref{zofzeta}), takes values in the whole real axis $(-\infty, +\infty)$. The parameter $r_S$  defines the Schwarzschild mass $r_S=2M$. The length $l_\epsilon$ is related to the scale $\epsilon$, as 
$\epsilon=-2l_\epsilon^2$. The $H(x)$ function is given by
\be
H(x)=-\frac{1}{\delta_c}+\frac{1}{2}\sqrt{z^4(x)-1}[f_{3/4}(x)+f_{7/4}(x)],
\ee
with
\be
f_\lambda(x)={_2F}_1[1/2,\lambda,3/2,1-z^4(x)]
\ee
being a hypergeometric function, and $\delta_c\approx 0.572069$ is an integration constant needed to find the correct far away asymptotic behavior.

The different parameters appearing in the line element \eqref{metric} can be rewritten as functions of the dimensionless parameters $N_q \equiv q/e$, with $e$ representing the proton charge, and the charge-to-mass ratio $\delta_1$, defined in Eq.~\eqref{defdelta1}. Let us explicitly write these relations, as 
\begin{equation}
q=eN_q \ , \ r_q=2l_P N_q/N_c \ , \ r_S=\frac{r_c^3}{2\delta_1 l_\epsilon^2} \ ,
\end{equation}
where $l_P$ is the Planck length and $N_c\equiv \sqrt{2/\alpha_{em}}\approx 16.55$ is a critical number of charges, which represents the transition from BH ($N_q>N_c$) to naked wormhole ($N_q<N_c$). In the definition of $N_{c}$, $\alpha_{em}$ is the fine structure constant. These definitions show how the line element \eqref{metric} is totally specified by the two dimensionless parameters $(\delta_1, N_q)$ plus the scales $l_\epsilon$  and $l_P$. This family of metrics leads to qualitatively different kinds of spacetimes, depending on the values of $\delta_1$ and $N_q$. The classification goes as follows: 
\begin{itemize}
\item {\bf Schwarzschild like solutions}: Characterized by $\delta_1 < \delta_c$, they possess an event horizon 
(on each side of the wormhole) for all values of $N_q$.
\item {\bf Reissner-Nordstr\"om like solutions}: With $\delta_1 > \delta_c$, they may exhibit two, one (degenerate), or no horizons (on each side of the wormhole), like in the usual Reissner-Nordstr\"om (RN) solution of GR. 
\item {\bf Regular solutions}: With $\delta_1 = \delta_c$, if $N_q > N_c$, one finds one horizon on each side
of the wormhole (similar to the Schwarzschild case). If $N_q = N_c$, the two symmetric horizons 
meet at the wormhole throat, $r = r_c$ (or $ x= 0$). For $N_q < N_c$  the horizons disappear yielding
a wormhole that connects two asymptotically Minkowskian universes. We will refer to these solutions as BH remnants, as they are continuously connected with BH configurations. The existence of
such remnants, which may arise at the end of BH evaporation or due to large density fluctuations in the early Universe \cite{Hawk}, might be of special relevance for the understanding
of the information loss problem \cite{Chen:2014jwq} and may also have potential observational consequences \cite{Lobo:2013prg}.  It is important to note that when the charge-to-mass ratio $\delta_1$ is set to the value $\delta_c$, the mass spectrum of the solutions is completely determined by the charge parameter $N_q$, through the relation \cite{Olmo:2013gqa}
\begin{equation}
M=m_P \left(\frac{N_q}{N_c}\right)^{3/2}\left(\frac{l_P}{l_\epsilon}\right)^{1/2} n_{BI} \ ,
\end{equation}
where $n_{BI}= \pi^{3/2}/(3\Gamma[3/4]^2)\approx 1.23605$. Up to a $\sqrt{2}$ numerical factor, this mass/energy expression is identical to the one found for point charges in the Born-Infeld electromagnetic theory. 
\end{itemize}
All the above cases rapidly tend to the standard GR solutions just a few $r_c$ units away from the wormhole throat at $r_c$. This puts forward the idea that the wormhole structure and deviations from GR arise nonperturbatively. It is worth noting that the area of the wormhole, $\mathfrak{A}=4\pi r_c^2=(8\pi l_\epsilon l_P/N_c) N_q$, grows linearly with the number of charges as if it was quantized. 

\section{Null rays and scalar waves}
\label{sec:Abs}

We now shift our attention to the properties of geodesics and scalar waves propagation in the spacetime defined by the line element (\ref{metric}). Since it represents a  static and spherically symmetric geometry, we have two Killing fields, $\xi_1=\partial_t$ and $\xi_2=\partial_\varphi$, which satisfy $u^a\nabla_a (g_{bc}u^b\xi_i^c)=0$, along metric geodesics with tangent vector $u^a$ \cite{Wald}. 
Due to spherical symmetry, we may restrict our attention to geodesics at the equatorial plane ($\theta=\pi/2$), without loss of generality.  
The two Killing vectors $(\partial_{t},\partial_{\phi})$ give the following conserved quantities along equatorial geodesics~\footnote{Note that in metric-affine gravity the geodesics of the connection are not those of the metric. Nonetheless, since the scalar field equations are not sensitive to the affine 
structure, they will follow metric geodesics in the eikonal limit. For geodesics of the connection, $u^a\nabla_a (g_{bc}u^b\xi^c)=0$ is not generally satisfied by a Killing field.} 
\be\label{Killing}
\begin{split}
&E=-A\dot{t} , \\
&L=r^2(x)\dot{\varphi},
\end{split}
\ee
where the overdot means derivative with respect to the affine parameter of the corresponding geodesic.

\subsection{Capture of null geodesics}

For metric geodesics we also have conservation of the norm of their tangent vector, $g_{bc}u^bu^c=-k$, where $k=0$ for null geodesics and $k=1$ for (affine-parametrized) timelike geodesics. Using Eqs.~\eqref{metric} and \eqref{Killing} we can thus write

\be\label{eqforzeta}
\frac{1}{2}m(x)\dot{x}^2+V_{eff}(x)=E^2,
\ee
with the definitions
\be
\label{pot}
V_{\rm eff}(x)\equiv A(x)\left(\frac{L^2}{r^2(x)}+k\right) ,
\ee
and
\be
\label{mtilde}
m(x)\equiv 4/\mathcal{Z}^2_+(x) .
\ee

Equation~\eqref{eqforzeta} is similar to that of a particle of mass $m(x)$ and energy $E^2$ in a central effective potential $V_{eff}$. 
\begin{figure}[h]
\includegraphics[width=\columnwidth]
{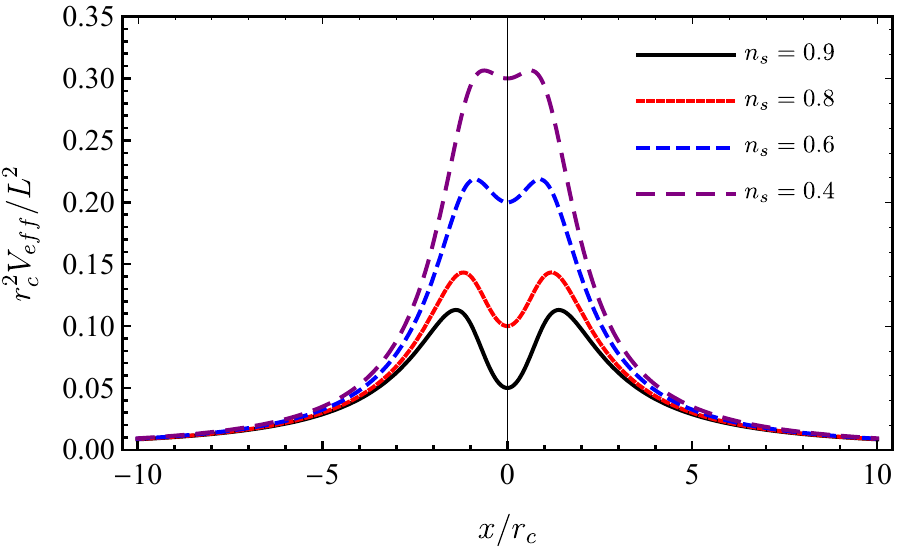}
\caption{Effective potential $V_{\rm eff}(x)$, given by Eq.~\eqref{pot} and normalized by $r_c^2/L^2$, for null geodesics ($k=0$) in BH remnant spacetimes. Here $n_s = N_q / N_c$. Note that for values of $n_s\approx 1$ we have a more pronounced potential well at $x=0$, and the well disappears as $n_s\rightarrow 0$ (or $N_q\rightarrow 0$).}
\label{fig:claspotentials}
\end{figure}

As we can see in Fig.~\ref{fig:claspotentials}, the effective potential associated to null geodesics presents a well at the wormhole throat, with one maxima on each side, defining two unstable photonspheres. The minimum of the potential, at the wormhole throat, is related to a stable photonsphere. We can see that the depth of the potential well increases as the normalized number of charges $n_s$ increases. Intuitively, concerning scalar waves,  this is telling us that the wormholes will be more absorptive the more charged they are, because their area grows linearly with the charge. Moreover, from the presence of a potential well, we can anticipate the existence of quasibound modes around the throat in the 
wave regime. Indeed, the absorption spectrum of scalar waves, 
computed in Subsec.~\ref{sec:TM}, shows the existence of these modes.\\   

The light-rings are circular geodesics such that $\dot{x}=0$ and $\ddot{x}=0$, 
corresponding to the maxima of the effective potential $V_{\rm eff}(x)$, given by Eq.~\eqref{pot}, 
with $k=0$.

Null geodesics impinging from infinity, which reach and stay at the maximum of the potential are called \textit{critical},
and they are characterized  by $V_{\rm eff}(x_{\rm max})=E^2$. This relation fixes their impact parameter, $b \equiv L/E$, to be 
\begin{equation}\label{Bcritclas}
b_c=\sqrt{\frac{L^2}{V_{\rm eff, max}}}=\frac{r_{\rm max}}
{\sqrt{A_{\rm max}}},
\end{equation} 
where the subindex $\rm max$ denotes evaluation of the corresponding function at $x_{\rm max}$. The critical impact parameter is related to the frequency of the unstable circular null geodesic by
\be
\Omega_l=b_c^{-1}.
\label{eq:frequency_null}
\ee
Null geodesics with $b>b_c$ are scattered by the BH remnant and stay 
in Region I (defined in Subsec.~\ref{scalar_absorption}), 
whereas those with $b<b_c$ overcome $V_{\rm  eff,max}$ and cross the wormhole throat 
to Region II. 
The classical absorption cross section for BH remnants is then given by

\begin{equation}\label{ClasAbsCrossBHremnants}
\sigma_{c} = \pi b_c^2 = \frac{\pi r_{\rm max}^2}{A_{\rm max}}.
\end{equation}

\begin{figure}[H]
\includegraphics[width=\columnwidth]{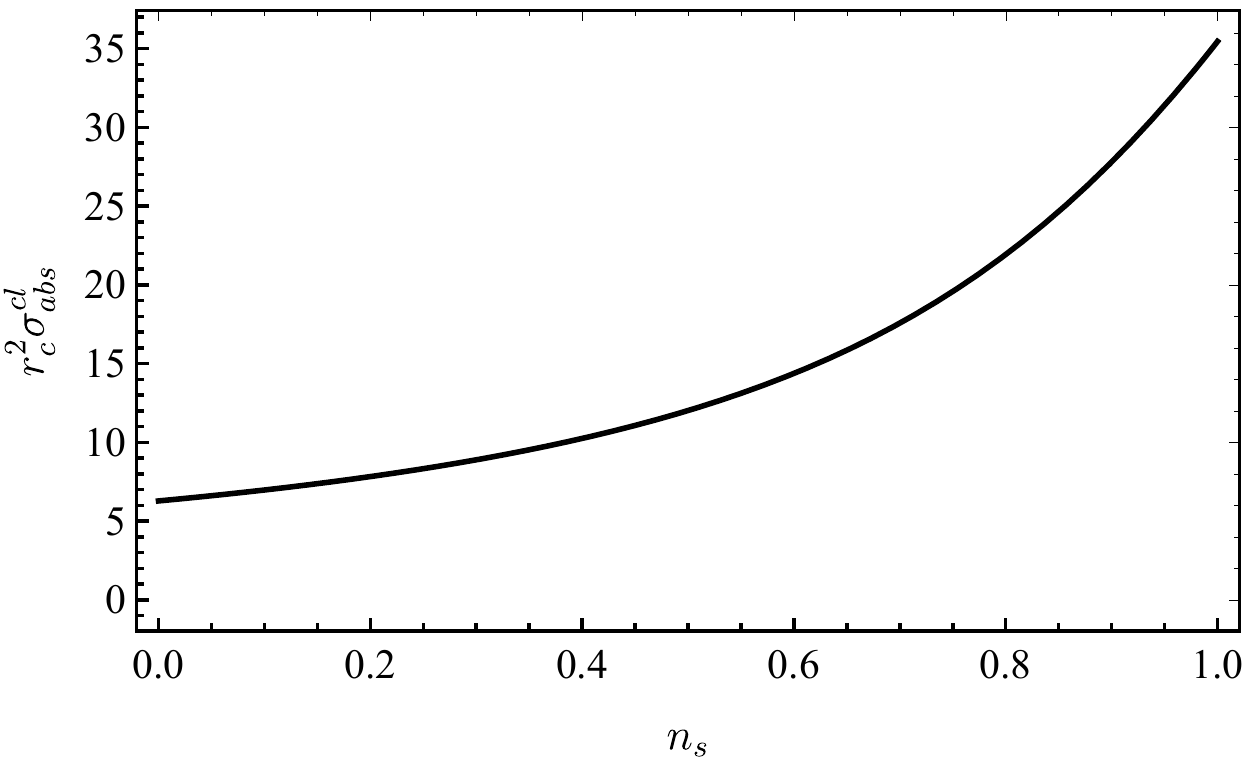}
\caption{Null geodesics absorption cross section of BH remnants for different values of $n_s$. Recall that BH remnant solutions are characterized by $\delta_1=\delta_c$ and  $n_s=N_q/N_c\in (0,1)$.}
\label{fig:ClassAbsCros}
\end{figure}

Despite that in our model it is not possible to solve $V'_{\rm eff, max}=0$ analytically, it is always possible to find  $ x_{max}$ through a numerical approach.  In Fig.~\ref{fig:ClassAbsCros} we present a plot of the total absorption cross section for null rays absorbed by a naked wormhole as a function of $n_s$. We can see that the absorption cross section increases monotonically with the (normalized) number of charges. Therefore, for an observer at infinity, $n_s$ can be regarded as an effective dissipative coefficient. As it will be seen later, the behavior of $n_s$ as a dissipative coefficient also holds for the absorption of scalar waves.

\subsection{Absorption of massless scalar waves}
\label{scalar_absorption}

As it is well known, the analysis of null geodesics is associated to the high-frequency limit (geometrical optics approximation) of planar massless waves scattering~\cite{Casti2005, Macedo2014}. The geodesic analysis, however, is not sensitive to the full range of phenomena that waves can experience, providing incomplete information about the absorption and scattering spectra, as well as the modal structure of the spacetime. These characteristics are also strongly dependent on the spin of the waves considered~\cite{chm, och2011, bodc, ldc}. As a first approach to this problem, we consider massless scalar waves, which provide interesting insights on the features of the spacetime beyond the geodesic approximation~\cite{Olmo:2015bya, bc2016, lbc2017} . \\

Let us consider a minimally coupled scalar field described by the action
\be
S_{\Phi}=\alpha_\Phi\int d^4x\sqrt{-g}\l[\frac{1}{2}\nabla_a\Phi\nabla^a\Phi+W(\Phi)\r],
\label{eq:scalaraction}
\ee
where $\Phi$ is the scalar field, $\alpha_\Phi$ is a coupling constant, and $W(\Phi)$ represents a self-interacting potential. We will assume a massless scalar field such that $W=0$. 
We can obtain the Klein-Gordon equation by extremizing the action \eqref{eq:scalaraction} with respect to $\Phi$, leading to
\be
\nabla_a\nabla^a\Phi=\frac{1}{\sqrt{-g}}\partial_a\l[\sqrt{-g}\partial^{a}\Phi\r]=0.
\label{eq:scalarwave}
\ee
The scalar field also enters into the modified Einstein equations through its stress-energy tensor. However, since we are considering the scalar field as a perturbation propagating in the spacetime, we shall neglect all backreaction terms, which are of ${\cal O}(\Phi^2)$, at least. We will thus focus on solving Eq.~\eqref{eq:scalarwave} in the background described by the line element~(\ref{metric}), with appropriate boundary conditions.

Once that the spacetime is spherically symmetric, we use separation of variables to decompose the field as
\be
\Phi=\frac{\varphi(t,x)}{r(x)}Y_{\ell m}(\theta,\phi),
\label{eq:scalardecom}
\ee
where $Y_{\ell m}(\theta,\phi)$ are the scalar spherical harmonics. 
Plugging Eq.~\eqref{eq:scalardecom} into Eq.~\eqref{eq:scalarwave}, after elementary manipulations we obtain the one-dimensional wave equation
\be
\l(\frac{\p^2}{\p r_\star^2}-\frac{\p^2}{\p t^2}-V_\varphi(r_\star)\r)\varphi(x, t)=0,
\label{eq:timeeq}
\ee
with the effective potential $V_\varphi$ given by
\be
V_\varphi(r_\star)=\frac{A\ell(\ell+1)}{r^2}+\frac{d^2r}{d{r^2_\star}} \ ,
\label{Vvarphi}
\ee
where $r_\star$ represents a tortoise-like coordinate, defined as 
\be
{dr_{\star}}\equiv \frac{dx}{A\,\mathcal{Z}_+} \ .
\ee
It is easy to check that $\frac{d^2r}{d{r^2_\star}}$ can be written as
\begin{equation}
\frac{d^2r}{d{r^2_\star}}= A\mathcal{Z}_+\frac{d}{dx}\lr{A\mathcal{Z}_+\frac{dr}{dx}}\, .
\end{equation}
Equation~\eqref{eq:timeeq} 
can be reduced to an ordinary differential equation by decomposing the function $\varphi(t,x)$ as $\varphi(t,x)=\varphi(x) e^{-i\omega t}$, leading to
\be
\l(\frac{d^2}{d r_\star^2}+\omega^2-V_\varphi(r_\star)\r)\varphi(x)=0.
\label{eq:freqeq}
\ee

\begin{figure*}
\includegraphics[width=\columnwidth]{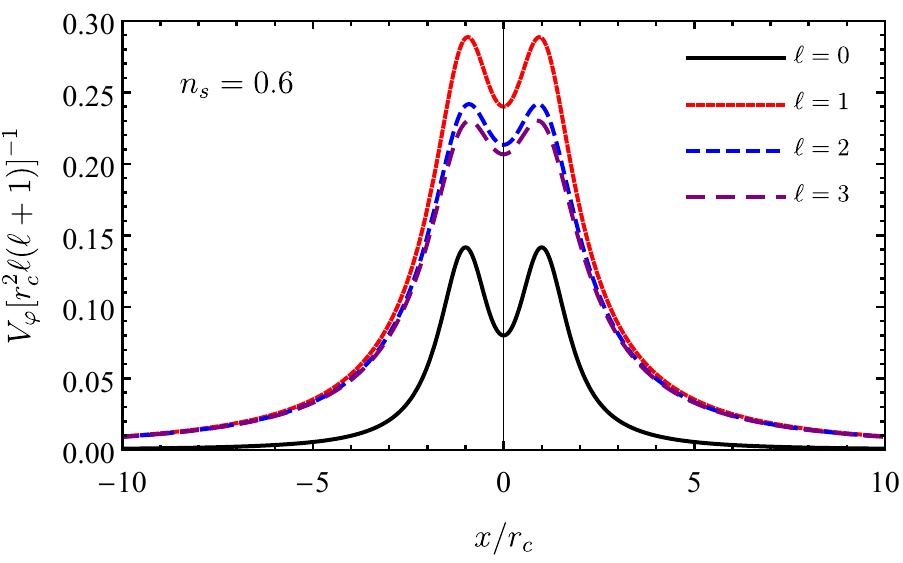}
\includegraphics[width=\columnwidth]{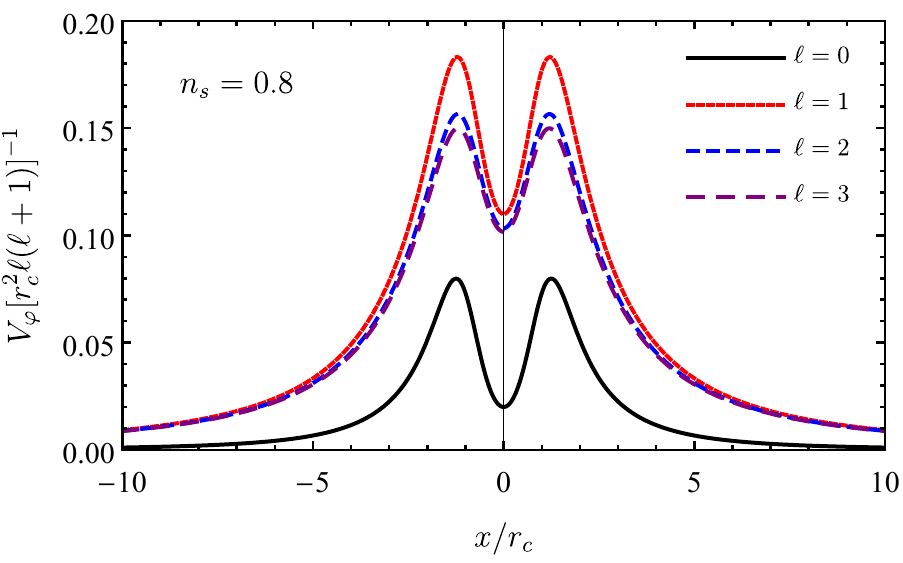}
\includegraphics[width=\columnwidth]{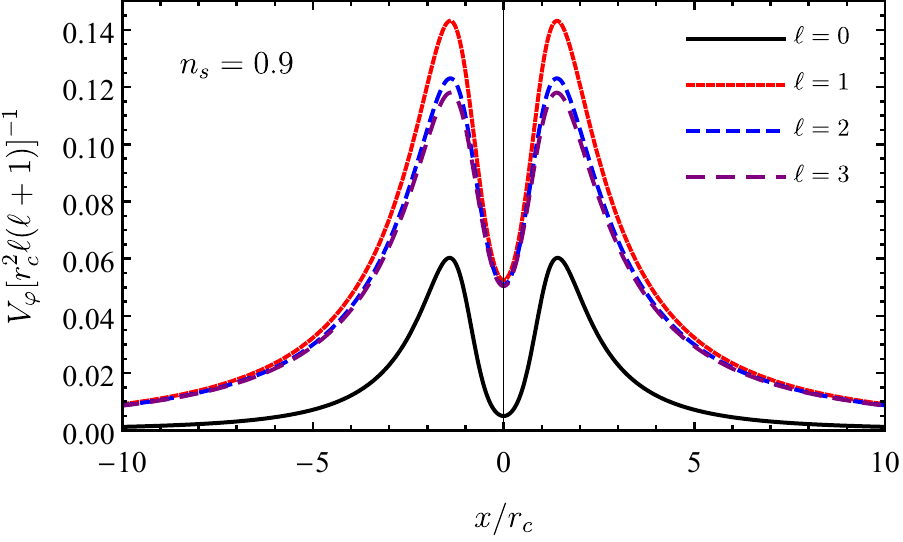}
\includegraphics[width=\columnwidth]{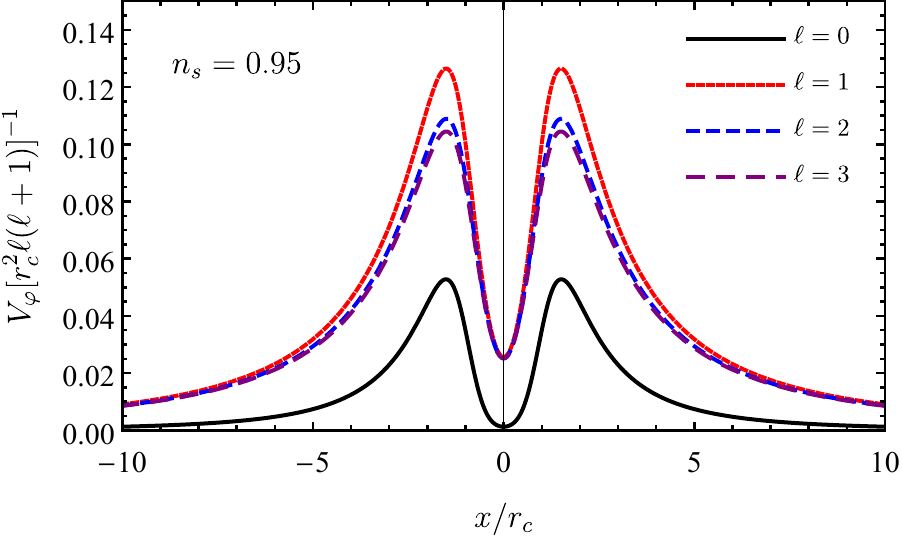}\\
\caption{Effective scalar potential $V_\varphi$ [given by Eq.~\eqref{Vvarphi}] for BH remnants. 
The plots with $\ell>1$ are normalized by $\ell(\ell+1)$, for better visualization. 
We note the presence of a well centered at $x=0$, which gets deeper as the limit $n=1$ is approached, and is shallower for higher multipoles. We also note the similarity of $V_\varphi$ with the potential obtained in the null geodesic analysis, $V_{\rm eff}$ [given by Eq.~\eqref{pot}], plotted in Fig.~\ref{fig:claspotentials}.}%
\label{fig:potentialsphi}%
\end{figure*}

In Fig.~\ref{fig:potentialsphi} we show the effective scalar potential $V_\varphi$, given by Eq.~\eqref{Vvarphi}, for different choices of $n_s$. For BH remnants, we have that a potential well may appear at $r=r_c$, showing different features from the BH case. The potential is consistent with the one from the geodesic analysis, 
given by Eq.~(\ref{pot}),
and the potential well is smaller for small values of $n_s$.

\begin{figure}
\includegraphics[width=1\columnwidth]{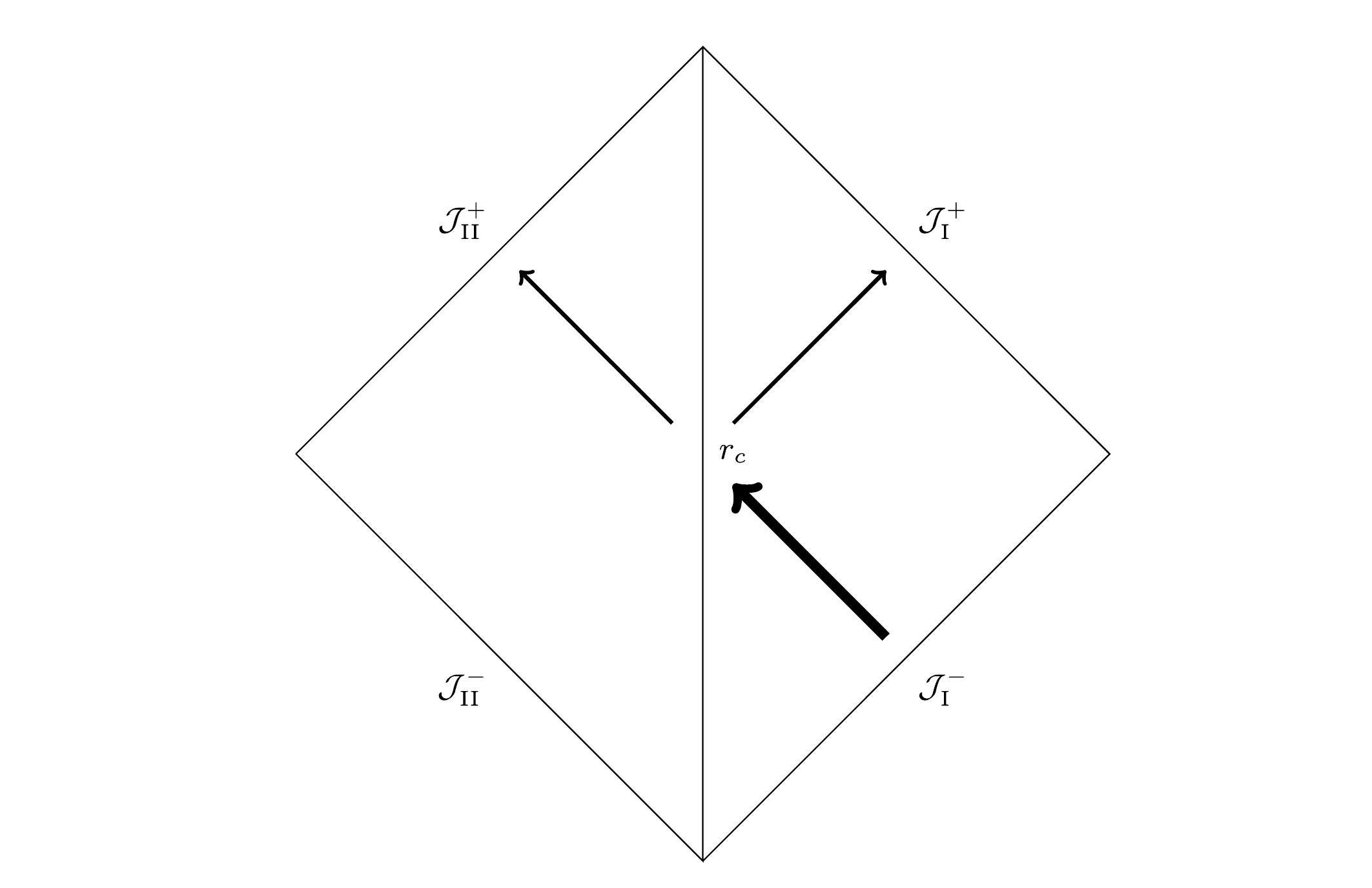}
\caption{Penrose diagram for a BH remnant configuration ($\delta_1=\delta_c$ and $N_q<N_c$). The wormhole is represented by the vertical timelike trajectory denoted by $r_c$. The triangular sectors on each side represent two asymptotically Minkowskian universes. The arrows represent the scattering of the massless scalar waves.}
\label{fig:PenDiagNWH}
\end{figure}

Proper boundary conditions should be supplemented to Eq.~\eqref{eq:freqeq}. The Penrose diagram of remnant configurations and the corresponding illustration of the scattering problem is depicted in Fig. \ref{fig:PenDiagNWH}. The right and left hand sides of the diagram are identified as Regions I and II, respectively. We are interested in planar waves incoming from past (null) infinity on the bottom right part of the diagram, ${\cal J}_{\rm I}^-$, being reflected to ${\cal J}_{\rm I}^+$ and transmitted to ${\cal J}_{\rm {II}}^+$. 
In Region I, asymptotically ($r \to +\infty$), we have
\be
\varphi_{\rm R_I}(x)\approx{\cal A}_{\ell m} e^{-i \omega r_\star}+{\cal R}_{\ell m} e^{i \omega r_\star},
\label{eq:phiari}
\ee
where ${\cal A}_{\ell m}$ is the amplitude of the incoming wave and ${\cal R}_{\ell m}$ the amplitude of the reflected one. To write Eq.~(\ref{eq:phiari}) we have used the fact that the potential vanishes asymptotically. 
The wave coming from ${\cal J}_{\rm I}^-$ is scattered by the compact object, leading to a phase difference between ${\cal A}_{\ell m}$ and ${\cal R}_{\ell m}$. 
The compact object can also partially absorb the wave, resulting in a difference in the moduli of the amplitude of the incoming and reflected wave. In the case of BHs, the absorption is associated to a purely ingoing wave into the horizon. 
For wormholes, which is the case of the BH remnant treated here, we identify the absorption with the part of the wave that is transmitted through the throat, emerging in the other universe (Region II). 

%
%
%

To describe the scattering phenomenology, we have to compute the phase-shift  $\delta_{\omega \ell}$, which is related to the reflection coefficient by
\be
e^{2i\delta_{\omega \ell}}=(-1)^{\ell+1}\frac{{\cal R}_{\omega \ell}}{{\cal A}_{\omega \ell}}.
\ee
In general, the phase-shift is complex whenever $|{\cal R}_{\omega \ell}|\neq|{\cal A}_{\omega \ell}|$, i.e., when there is a dissipation in the system. 

The absorption cross section is given by~\cite{Futterman:1988ni}
\be
\sigma=\sum_{\ell=0}^{\infty}\sigma_{\ell},
\ee
where $\sigma_l$ is the partial absorption cross section, given by
\begin{align}
\sigma_l&=\frac{\pi}{\omega^2}(2\ell+1)	\left[1-e^{4 {\rm Im}(\delta_{\omega \ell})}\right]\\
&=\frac{\pi}{\omega^2}(2\ell+1) \Gamma_{\omega \ell}\label{sigmagamma},
\end{align}
with
\be
\Gamma_{\omega \ell}=1-\l|\frac{{\cal R}_{\omega \ell}}{{\cal A}_{\omega \ell}}\r|^2
\ee
being the transmission coefficients. To compute the reflection coefficient, we must impose that the boundary conditions in the asymptotic limit of Regions I and II are satisfied, resulting in equations for the amplitude of the wave in those limits. This can be done by analytical approximations of the wave function or by numerically integrating it from the asymptotic limit of Region II to the asymptotic limit of Region I, and comparing the result with the asymptotic form given by Eq.~\eqref{eq:phiari}.\\

\subsection{Trapped modes}
\label{sec:TM}

Due to the shape of the potential, quasibound states can exist, associated to the potential well located at $r=r_c$. These quasibound states are similar to the trapped modes present in ultracompact stars~\cite{Kokkotas:1999bd}, and in the eikonal limit they are related to the stable null-geodesics existing at $r=r_c$~\cite{Cardoso:2014sna}. The modes are complex, having small imaginary part due to the tunneling to the asymptotic regions of the spacetime. They are determined by the boundary conditions
\be
\varphi=\left\{
\begin{array}{ll}
	e^{-i\omega r_\star},& x\to-\infty,\\
	e^{i\omega r_\star},& x\to\infty,
\end{array}\right.
\ee
which generates an eigenvalue problem for the frequency $\omega$. 
The existence of trapped modes in the BH remnant case is a crucial difference from the BH spacetime, where the imaginary part is associated to the timescale of the unstable null geodesic~\cite{Cardoso:2008bp}. 
The quasibound modes generate a signature in the absorption spectrum, leading to narrow spectral lines in it. 
In fact, this signature has been also found in weakly dissipative ultracompact stars~\cite{Macedo:2018yoi}, 
where it was shown that the trapped modes give rise to structures similar to the Breit-Wigner resonances in nuclei scattering. The position and the structure of the spectral lines depend on the nature of the compact object and may, therefore, be used to distinguish it from other cases.

\begin{figure}
\includegraphics[width=\columnwidth]{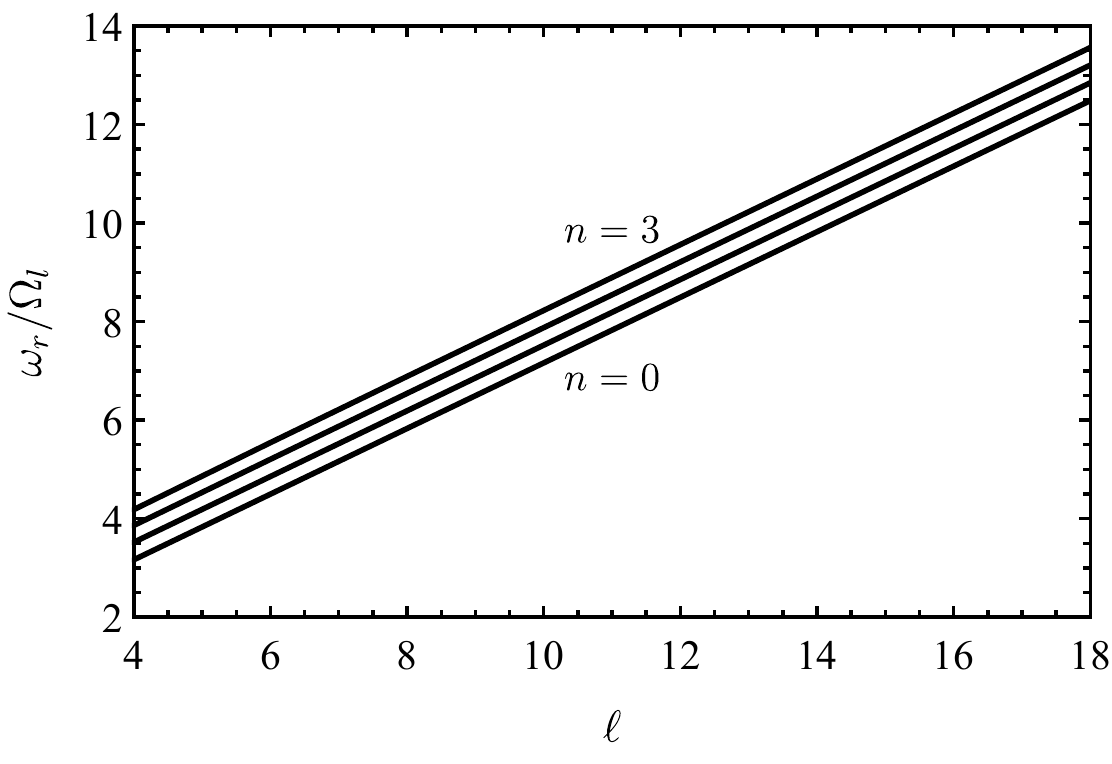}%
\caption{Real part of the fundamental ($n=0$) and first three overtones ($n=1,2,3$) frequencies of the trapped modes, obtained through Eq.~(\ref{eq:quantization}), as a function of $\ell$, for the case $n_s=0.9$.}%
\label{omega-r}%
\end{figure}

In the eikonal limit, the real part of the trapped modes $\omega_r$ can be found through the Born-Sommerfeld quantization rule~\cite{Gurvitz:1988zz}
\be
\int_{r_{\star a}}^{r_{\star b}}dr_\star\sqrt{\omega_r^2-V_{\varphi}(r_\star)}= \pi(n+1/2),
\label{eq:quantization}
\ee
where ${\omega_r}^2<V_{\varphi}$, $n$ is a positive integer, and  $r_{\star a,b}$ are the inner turning points, defined through ${\omega_r}^2-V_{\varphi} =0$. As previously mentioned, the imaginary part of these modes is usually very small, what leads to resonant narrow peaks in the transmission coefficient. We shall use Eq.~\eqref{eq:quantization} to describe the position of the resonant peaks. From Eq.~\eqref{eq:quantization}, we obtain that the modes fit as~\cite{Cardoso:2014sna}
\be
\omega_r\sim a\ell+b,
\label{eq:bound_cond}
\ee
where $b$ is a constant (cf. Fig.~\ref{omega-r}), that depends on the overtone number, and
\be
a=\lim_{x\to 0}\frac{A(x)^{1/2}}{r(x)},
\ee
being the angular frequency of the stable null geodesic. The above result tells us that the frequency of the trapped modes is evenly spaced with the overtone number. Such characteristic will generate interesting patterns in the transmission coefficient.

In addition to the trapped modes, an approximation based on the Breit-Wigner expression for nuclei scattering can be used to describe the absorption cross section. Essentially, when $\omega\approx \omega_r$, we have~\cite{Macedo:2018yoi}
\be
\left.\Gamma_{\omega \ell}\right|_{\omega\approx\omega_r}\propto\frac{1}{(\omega-\omega_r)^2+\omega_i^2},
\ee
where $\omega_i$ is the imaginary part of the mode. We can see that the transmission factor peaks at $\omega=\omega_r$ with a height that depends on the imaginary part of the mode. Conversely, the above expression can also be used to extract the trapped modes frequencies from the computation of the transmission factor.

\section{Numerical Results}
\label{sec:NR}

Let us now present the numerical results for the absorption cross section of planar massless scalar waves by BH remnants. The absorption properties are intrinsically related to the geodesic quantities, as noted before. Therefore we choose to normalize the absorption cross section by its corresponding classical limit. Such normalization brings our results closer to observational quantities, and it also makes easier to compare them with those obtained for BHs within GR.

\begin{figure*}
\includegraphics[width=\columnwidth]{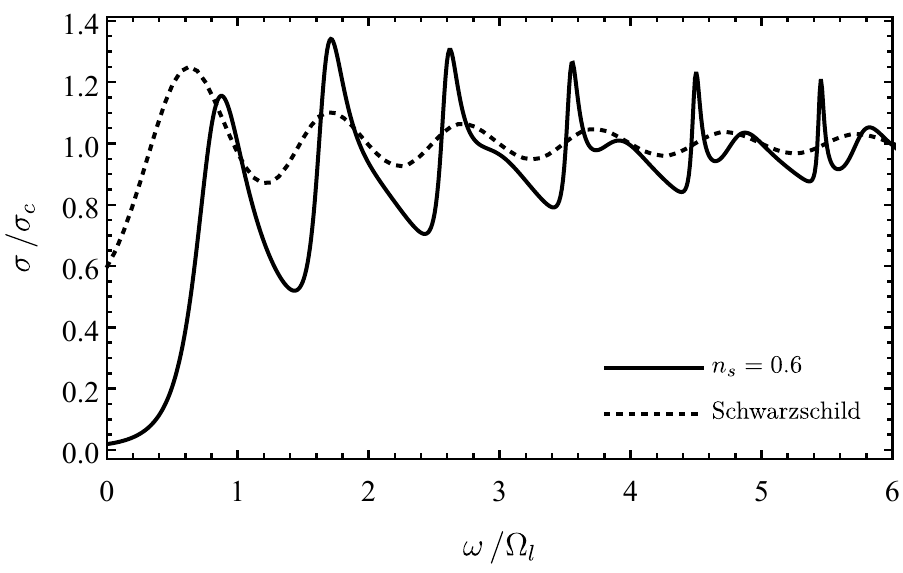}\includegraphics[width=\columnwidth]{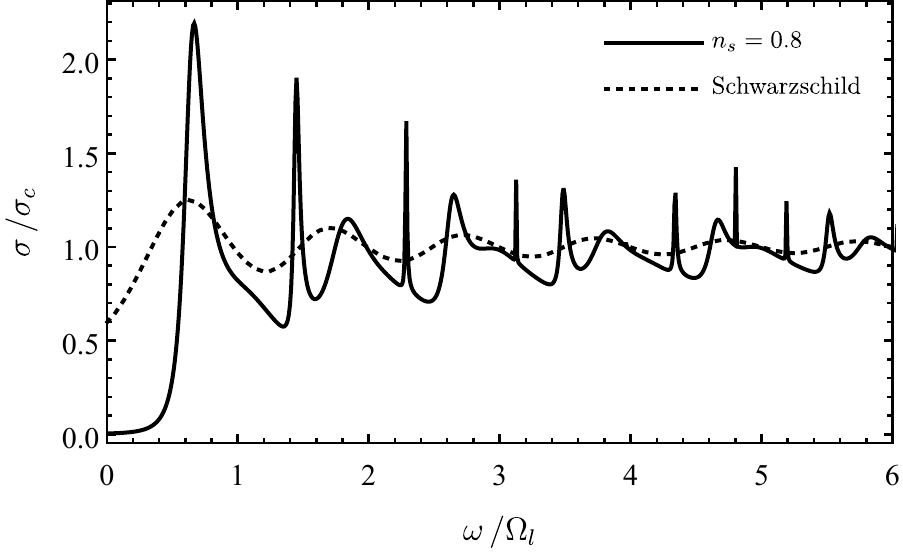}\\
\includegraphics[width=\columnwidth]{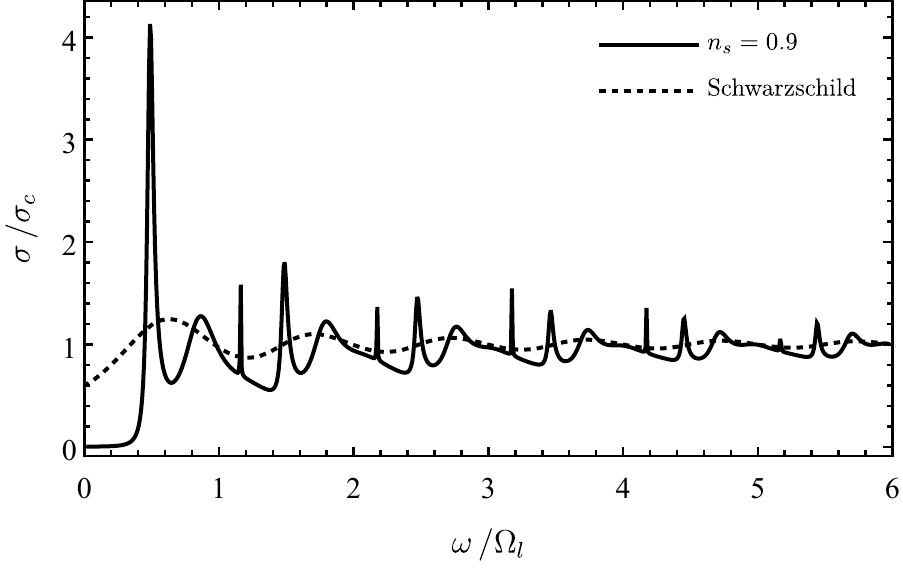}\includegraphics[width=\columnwidth]{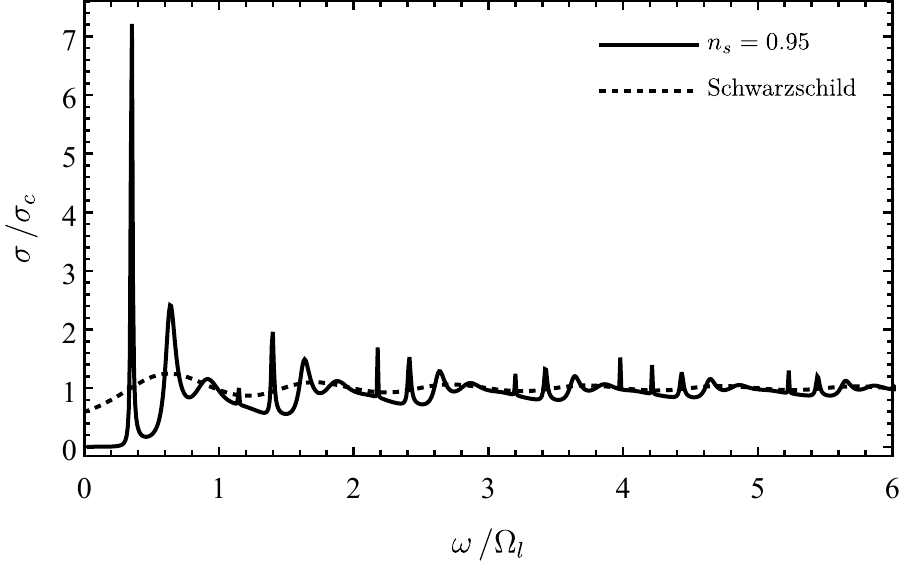}
\caption{Scalar absorption cross section of BH remnants for different values of $n_s$. The absorption cross section is normalized by the classical cross section and the frequency by the light-ring frequency. Narrow peaks arise when trapped modes exist in the potential well. The dotted lines correspond to the Schwarzschild BH case.}%
\label{fig:ScalarAbsCross}%
\end{figure*}

\begin{figure*}
\includegraphics[width=\columnwidth]{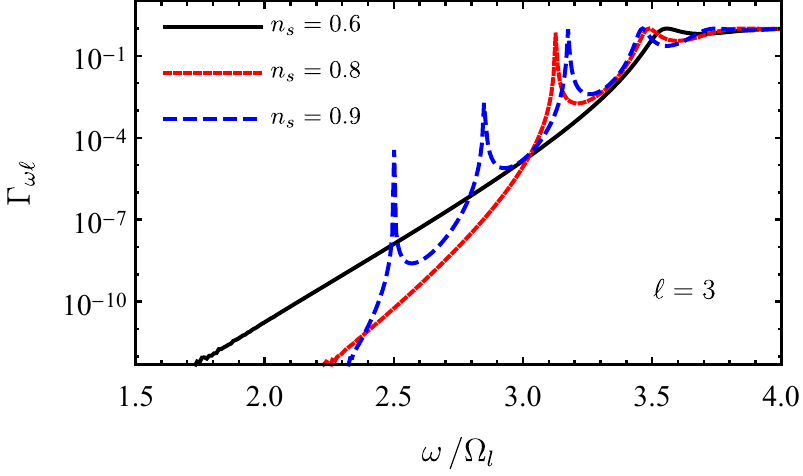}\includegraphics[width=\columnwidth]{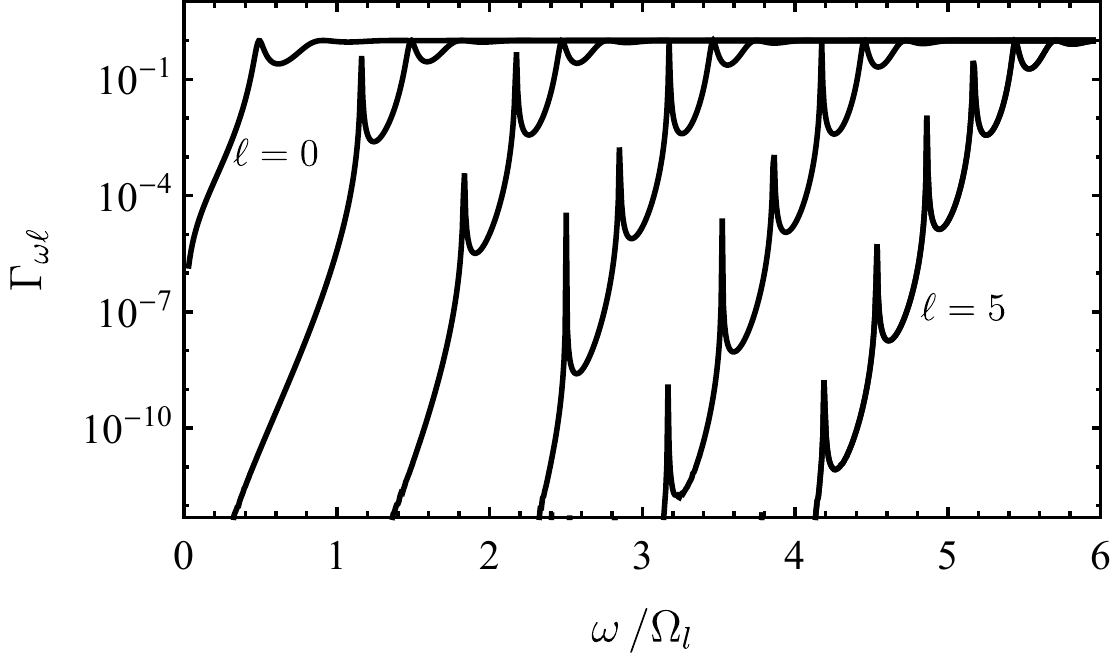}%
\caption{Representative cases for the transmission coefficient in BH remnants. \textit{Left panel:} As noted in the behavior of the effective scalar potential $V_\varphi$, trapped modes are more likely to arise for $n_s\approx 1$, and this feature impacts the transmission coefficients, generating resonant peaks. \textit{Right panel:} In addition to the $n_s$ dependence, more resonant peaks appear for higher values of $\ell$, as illustrated here for the $n_s=0.9$ case.}%
\label{fig:TransfCoeff}%
\end{figure*}

In Fig.~\ref{fig:ScalarAbsCross} we plot the absorption cross section for massless planar scalar fields as a function of the frequency, which we normalized by the light-ring frequency value $\Omega_l$ [given by Eq.~\eqref{eq:frequency_null}]. The absorption cross section is normalized by its classical counterpart, so that the plots in Fig.~\ref{fig:ScalarAbsCross} tend to the unity in the high-frequency regime. We note that the absorption in the low-frequency regime is different from the Schwarzschild BH result, showing a Breit-Wigner type resonant behavior for some given frequencies, what indicates the presence of trapped modes in the potential well around the wormhole throat. This result is analog to recent findings regarding ClePhOs' absorption spectrum, reported in Ref.~\cite{Macedo:2018yoi}. 
We note that 
the (normalized) number of charges $n_s$ is analog to the absorption parameter $\mathcal{K}$ of Ref.~\cite{Macedo:2018yoi}. Fig.~\ref{fig:TransfCoeff} is a plot of the transmission coefficient of BH remnants as a function of the frequency. 
From the left panel of Fig.~\ref{fig:TransfCoeff}, it can be seen that for a fixed multipole $\ell$, the number of peaks increase as $n_s$ approaches the unity. Moreover, the number of peaks for a fixed value of $n_s$ increases as we increase the multipole number $\ell$, as it can be seen in the right panel of Fig.~\ref{fig:TransfCoeff}. These different peaks enter at different frequency regimes, as it can be seen in the absorption plots (cf. Fig.~\ref{fig:ScalarAbsCross}). For a peak to be pronounced in the absorption spectrum it has to have a frequency high enough to penetrate the potential barrier, i.e., $\omega^2\sim V_{\rm eff,max}$. We note from Eq.~(\ref{sigmagamma}) that the absorption cross section contains a multiplicative factor of $\omega^{-2}$.

\section{Phenomenological implications} \label{sec:PhenoImp}

The results presented in the two previous sections indicate that the family of objects considered in this work has similar absorptive properties as ClePhOs. 
The existence of a Breit-Wigner like structure is due to the fact that the remnants, like ClePhOs, have an inner structure with two characteristic surfaces on which the waves can resonate.
Nonetheless, though the peaks in the absorption spectrum of dissipative star-like ClePhOs are similar to the case of the BH remnants explored here, the overall absorption is different, what may provide an observational discriminator for the existence of event horizons in different kinds of compact objects. We notice that at higher frequencies, the absorption by BH remnants tends to the Schwarzschild BH result, being equivalent to the capture cross section of null geodesics. This is not the case for weakly dissipative star-like ClePhOs, for which the high-frequency absorption cross section tends to $\sigma_c(1-|{\cal K}|^2)$, with $| {\cal K}| \approx 1$. Therefore, there is clearly a distinctive signature of star-like ClePhOs that allows to discriminate them from wormhole remnants.~\footnote{Recall that for $n_s=1$ our solutions develop a horizon, so that we are restricted to $0<n_s<1$.}

Another feature of these BH remnants that could be analysed in order to find observable discriminators from standard BHs and/or other ECOS is  their emission spectrum.  In this regard, note that the emission spectrum of ECOs will also have characteristic lines  described by $\Gamma_{\omega l}$ and, therefore, their emission spectrum will probably also be similar to that of ClePhOs, what further hinders their distinguishability within the ECO family. We also note that, since these features depend on the geometric properties of the objects, electromagnetic and gravitational perturbations may present similar characteristics.

Since this work is a first step in understanding the implications of BH remnants in the wave phenomenology regime, we have focused on the scalar case for simplicity. While this is important to get a grasp on more complex structures, it should be clear that an analysis of the full gravitational wave perturbations is needed to further quantify the physical phenomena explored in this paper. In this sense, we expect that a Breit-Wigner like spectrum will still be present for gravitational waves and will converge to the scalar field one in the high-frequency limit, where the geodesic approximation is valid. However, at lower frequencies it is difficult to anticipate quantitative results. In fact, since the spectrum of gravitational waves is directly related with the dynamics of the perturbations of the underlying theory, the fact that the background solutions presented here can be derived from (at least) two different gravity theories (see Sec.\ref{sec:framework} for details) indicates that, given a background, different  spectra are likely to arise for different theories. This diversity of potential results is well-known in cosmology, where different theories may yield identical background expansion histories but lead to completely different linear perturbation  evolution equations for structure growth  \cite{Silvestri:2013ne}. Something similar occurs in other scenarios, such as in scalar field theories, where the conditions to yield identical background solutions and linear perturbations are very special \cite{Bazeia:2017gub,Bazeia:2011hj}. Thus, a dedicated analysis of the perturbation equations for gravitational waves and their phenomenological implications should be carried out for each gravity theory that generates the line element (\ref{metric}).

Though such an analysis is not yet available for the RBG family of theories, some general conclusions can be extracted from the basic properties of these theories. In particular, given that Ricci-based gravity theories recover Einstein's equations in vacuum, the propagation of gravitational perturbations only involves two polarizations that travel at the speed of light, which is an important viability test for modified theories of gravity, especially after the simultaneous observation of gravitational and electromagnetic radiation from a neutron star merger \cite{Lombriser:2015sxa,Lombriser:2016yzn,Baker:2017hug,Sakstein:2017xjx,Creminelli:2017sry,Ezquiaga:2017ekz,Ezquiaga:2018btd}. Additionally, since the modified dynamics of RBGs manifests itself via nonlinearities induced by the matter fields, the coupling between gravitational and matter modes must be important, potentially leading to new observational features.

In this respect, since the remnants considered here have an electric charge, gravitational perturbations couple with electromagnetic ones generating new modes and more spectral lines in the absorption spectrum, which could be analyzed by extending the methods previously developed in the literature~ \cite{chm, och2011, Zerilli:1974ai, CHO:2009, CDHO:2014, CDHO:2015}. Therefore, ideally, one can potentially observe the gravitational sector through  its imprint on the electromagnetic one, and the new couplings generated by the modified dynamics could help to discriminate between GR and other theories. In this sense, we notice that the nonlinearities of RBG have only been studied in microscopic systems~\cite{Latorre:2017uve}, and astrophysical scenarios shall reveal new physical implications of these nonlinear couplings.  The study of these interesting features lies beyond the scope of this paper and will be left for future works. \\

\section{Discussion and final remarks} \label{sec:discusscf}

We have studied the absorption properties of ECOs which 
arise naturally in a wide class of metric-affine gravity models beyond GR, 
emerging from a wormhole structure in the vicinity of the classical singularity. 
These ECOs are continuously related with BH solutions and, therefore, could represent a feasible end state after Hawking evaporation. We have shown that their scalar absorption spectrum is characterized by 
Breit-Wigner-like resonances, whose frequencies
are related to the distance between the wormhole throat and its photonsphere.~\footnote{Taking the classical limit, the maximum in the effective potentials of Fig.~\ref{fig:potentialsphi} corresponds to the position of the photonsphere.} 

This study aims to be the first step in the characterization of the interactions between wormhole ECOs and matter fields, revealing that they present absorptive spectral features very similar to those of star-like ClePhOs \cite{Macedo:2018yoi}. The implications of such result are two-folded: (i) They allow to distinguish ECOs from standard GR BHs at the observational level, and could also be used in order to discriminate between the different modified gravity approaches that are studied today and do not predict the existence of ECOs;
 (ii)  They can be used to distinguish wormhole and star-like ClePhOs, since their absorption spectra have distinctive features, like the high-frequency limit.

Following the classification of the solutions provided in Sec.~\ref{sec:framework}, our study has focused on spherically symmetric compact objects with $\delta_1=\delta_c$ and $0<n_s<1$, for which there is no event horizon (which are called BH remnants). The mass spectrum of this set of solutions is bounded above by (approximately) the Planck mass \cite{Olmo:2013gqa}, limiting their astrophysical motivation.~\footnote{More massive alternative BH remnants are possible if nonlinear effects in the matter sector are taken into account \cite{Olmo:2013mla}.}  Nonetheless, our analysis paves the road to the study of the spectral properties of  other types of solutions with higher astrophysical relevance.  In particular, the solutions studied here are geodesically complete and possess bounded curvature scalars everywhere. But there exists another branch of solutions, with different charge-to-mass ratio, $\delta_1>\delta_c$, for which curvature scalars diverge at the wormhole throat, despite being geodesically complete as well. In this part of the spectrum we find what could be seen as naked divergences,~\footnote{The geodesic completeness of such naked objects would make cosmic censorship hypotheses unnecessary.} as opposed to naked singularities (which are geodesically incomplete). 
Since the interaction of matter fields with regions of extreme (even divergent) curvature is well defined in scenarios with wormholes (see Ref.~\cite{Olmo:2015dba} for a concrete example involving the model considered here and  Ref.~\cite{Giveon:2004yg} for different models in GR), it is important to evaluate in detail the observable impact that such curvature divergences might have on the absorption and emission spectra of such objects. This analysis is currently underway and is expected to shed some new light on the observational features of compact naked objects with curvature divergences. \\

\begin{acknowledgments}

The authors would like to acknowledge partial financial support from the European Union's Horizon 2020 research and innovation programme under the H2020-MSCA-RISE-2017 Grant No. FunFiCO-777740.
We also thank Conselho Nacional de Desenvolvimento Cient\'ifico e Tecnol\'ogico (CNPq) and Coordena\c{c}\~ao de Aperfei\c{c}oamento de Pessoal de N\'ivel Superior (CAPES) -- Finance Code 001, from Brazil, for partial financial support.
A. D.  and G. J.O. thank Universidade Federal do Par\'a for kind hospitality. A. D. is supported by a PhD contract of the program FPU 2015 (Spanish Ministry of Economy and Competitiveness) with reference FPU15/05406.
G. J. O. is funded by the Ramon y Cajal contract RYC-2013-13019 (Spain).  This work is supported by the Spanish projects FIS2014-57387-C3-1-P  and FIS2017-84440-C2-1-P (MINECO/FEDER, EU), the project SEJI/2017/042 (Generalitat Valenciana), the Consolider Program CPANPHY-1205388, and the Severo Ochoa grant SEV-2014-0398 (Spain).
\end{acknowledgments}


\begin{thebibliography}{9}
 \bibitem{Johnson-McDaniel:2018uvs} 
  N.~K.~Johnson-Mcdaniel, A.~Mukherjee, R.~Kashyap, P.~Ajith, W.~Del Pozzo, and S.~Vitale,
  Constraining black hole mimickers with gravitational wave observations,
  arXiv:1804.08026.
  
  \bibitem{Cunha:2017wao} 
  P.~V.~P.~Cunha, J.~A.~Font, C.~Herdeiro, E.~Radu, N.~Sanchis-Gual, and M.~Zilhão,
   Lensing and dynamics of ultracompact bosonic stars,
  Phys.\ Rev.\ D {\bf 96}, 104040 (2017).
  
\bibitem{Herdeiro:2017phl} 
  C.~A.~R.~Herdeiro and E.~Radu,
Dynamical Formation of Kerr Black Holes with Synchronized Hair: An Analytic Model,
  Phys.\ Rev.\ Lett.\  {\bf 119}, 261101 (2017).

\bibitem{Herdeiro:2014goa} 
  C.~A.~R.~Herdeiro and E.~Radu,
  Kerr black holes with scalar hair,
  Phys.\ Rev.\ Lett.\  {\bf 112}, 221101 (2014).

\bibitem{Liebling:2012fv} 
  S.~L.~Liebling and C.~Palenzuela,
  Dynamical Boson Stars,
  Living Rev.\ Rel.\  {\bf 15}, 6 (2012)
  [Living Rev.\ Rel.\  {\bf 20}, 5 (2017)].
  
  \bibitem{Cardoso:2016oxy}
  V.~Cardoso, S.~Hopper, C.~F.~B.~Macedo, C.~Palenzuela, and P.~Pani,
  Gravitational-wave signatures of exotic compact objects and of quantum corrections at the horizon scale,
  Phys.\ Rev.\ D {\bf 94},  084031 (2016).

  \bibitem{AbbotGW}
  B.~ P.~Abbott {\it et al.} (Virgo and LIGO Scientific Collaborations), 
  Gravitational Waves and Gamma-Rays from a Binary Neutron Star Merger: GW170817 and GRB 170817A,
  Astrophys. J. Lett. \textbf{848}, L13 (2017).

 \bibitem{AbbotPM}
  B.~ P.~Abbott {\it et al.} [(Virgo and LIGO Scientific Collaborations), 
  GW170817: Observation of Gravitational Waves from a Binary Neutron Star Inspiral,
   Phys. Rev. Lett. \textbf{119}, 141101 (2017).

  \bibitem{GBM:2017lvd}
  B.~P.~Abbott {\it et al.}, 
  Multi-messenger Observations of a Binary Neutron Star Merger,
  Astrophys.\ J.\  {\bf 848}, L12 (2017).

\bibitem{Abbott:2017dke}
  B.~P.~Abbott {\it et al.} (Virgo and LIGO Scientific Collaborations), 
  Search for Post-merger Gravitational Waves from the Remnant of the Binary Neutron Star Merger GW170817,
  Astrophys.\ J.\  {\bf 851},  L16 (2017).

  \bibitem{Abbott:2017gyy}
  B.~ P.~Abbott {\it et al.} (Virgo and LIGO Scientific Collaborations), 
  GW170608: Observation of a 19-solar-mass Binary Black Hole Coalescence,
  Astrophys.\ J.\  {\bf 851}, L35 (2017).

  \bibitem{Barack:2018yly}
  L.~Barack {\it et al.},
  Black holes, gravitational waves and fundamental physics: a roadmap, \textit{To appear in Classical Quantum Gravity, }
  arXiv:1806.05195.

\bibitem{Geroch:1968ut}
  R.~P.~Geroch,
  What is a singularity in general relativity?,
  Ann. Phys. (N. Y.) \  {\bf 48}, 526 (1968).

  \bibitem{Earman:1995fv}
  J.~Earman, 
  \emph{Bangs, crunches, whimpers, and shrieks: Singularities and acausalities in relativistic space-times}       (Oxford University Press, New York, 1995).
  
\bibitem{Wald:1997wa} 
  R.~M.~Wald,
  Gravitational collapse and cosmic censorship,
  In: B. Iyer and B. Bhawal (eds.), {\it Black holes, gravitational   	radiation and the universe 69-85}
  (Springer Netherlands, Heidelberg, 1999)

\bibitem{Berger:2002st} 
  B.~K.~Berger,
  Numerical approaches to spacetime singularities,
  Living Rev.\ Rel.\  {\bf 5}, 1 (2002).

\bibitem{Bardeen}
J. M. Bardeen, 
Non-singular general relativistic gravitational collapse,
{\it Proceedings of GR 5} 
(Tbilisi, USSR, 1968).

\bibitem{ABG98}
E. Ay\'on-Beato and A. Garc\'ia,
Regular Black Hole in General Relativity Coupled to Nonlinear Electrodynamics,
Phys. Rev. Lett. \textbf{80},  5056 (1998).

\bibitem{Bronnikov01}
K. A. Bronnikov, 
Regular magnetic black holes and monopoles from nonlinear electrodynamics,
Phys. Rev. D \textbf{63}, 044005 (2001).

\bibitem{Ansoldi:2008jw}
  S.~Ansoldi,
  Spherical black holes with regular center: A Review of existing models including a recent realization with Gaussian   sources,
  arXiv:0802.0330 [gr-qc].

\bibitem{DiazAlonso:2009ak}
  J.~Diaz-Alonso and D.~Rubiera-Garcia,
  Electrostatic spherically symmetric configurations in gravitating nonlinear electrodynamics,
  Phys.\ Rev.\ D {\bf 81}, 064021 (2010).
  
  \bibitem{DiazAlonso:2010eh}
  J.~Diaz-Alonso and D.~Rubiera-Garcia,
  Asymptotically anomalous black hole configurations in gravitating nonlinear electrodynamics,
  Phys.\ Rev.\ D {\bf 82}, 085024 (2010).
  
  \bibitem{DiazAlonso:2012mb}
  J.~Diaz-Alonso and D.~Rubiera-Garcia,
  Thermodynamic analysis of black hole solutions in gravitating nonlinear electrodynamics,
  Gen.\ Relativ.\ Gravit.\  {\bf 45}, 1901 (2013).
  
  \bibitem{Barcelo:2017lnx}
  C.~Barceló, R.~Carballo-Rubio, and L.~J.~Garay,
  Gravitational wave echoes from macroscopic quantum gravity effects,
  J. High Energy Phys. 05 (2017) 054.
  
  \bibitem{Fabbri:2005mw} 
  A.~Fabbri and J.~Navarro-Salas,
  \emph{Modeling black hole evaporation}
  (Imperial College Press, London, 2005).

\bibitem{Hayward:2005gi}
  S.~A.~Hayward,
  Formation and evaporation of regular black holes,
  Phys.\ Rev.\ Lett.\  {\bf 96}, 031103 (2006).
  
  \bibitem{Zhang:2014bea}
  Y.~Zhang, Y.~Zhu, L.~Modesto, and C.~Bambi,
  Can static regular black holes form from gravitational collapse?,
  Eur.\ Phys.\ J.\ C {\bf 75}, 96 (2015).
  
  \bibitem{Liu:2014kra}
  Y.~Liu, D.~Malafarina, L.~Modesto, and C.~Bambi,
  Singularity avoidance in quantum-inspired inhomogeneous dust collapse,
  Phys.\ Rev.\ D {\bf 90}, 044040 (2014).
  
  \bibitem{Malafarina:2017csn}
  D.~Malafarina,
  Classical collapse to black holes and quantum bounces: A review,
  Universe {\bf 3}, 48 (2017).

\bibitem{Rovelli:2014cta}
  C.~Rovelli and F.~Vidotto,
  Planck stars,
  Int.\ J.\ Mod.\ Phys.\ D {\bf 23}, 1442026 (2014).
  
  \bibitem{Spallucci:2017aod}
  E.~Spallucci and A.~Smailagic,
  Regular black holes from semi-classical down to Planckian size,
  Int.\ J.\ Mod.\ Phys.\ D {\bf 26}, 1730013 (2017).
  
  \bibitem{Abedi:2015yga}
  J.~Abedi and H.~Arfaei,
  Obstruction of black hole singularity by quantum field theory effects,
  J. High
Energy Phys. 03 (2016) 135.
  
  \bibitem{Ashtekar:2018lag} 
  A.~Ashtekar, J.~Olmedo, and P.~Singh,
  Quantum Transfiguration of Kruskal Black Holes,
  Phys.\ Rev.\ Lett.\  {\bf 121}, 241301 (2018).
  
  \bibitem{Bueno:2017hyj}
  P.~Bueno, P.~A.~Cano, F.~Goelen, T.~Hertog, and B.~Vercnocke,
  Echoes of Kerr-like wormholes,
  Phys.\ Rev.\ D {\bf 97}, 024040 (2018).

\bibitem{Olmo:2012nx} 
  G.~J.~Olmo and D.~Rubiera-Garcia,
  Reissner-Nordstr\"om black holes in extended Palatini theories,
  Phys.\ Rev.\ D {\bf 86}, 044014 (2012).

\bibitem{Olmo:2011np} 
  G.~J.~Olmo and D.~Rubiera-Garcia,
  Nonsingular black holes in quadratic Palatini gravity,
  Eur.\ Phys.\ J.\ C {\bf 72}, 2098 (2012).
  
 \bibitem{Olmo:2013gqa} 
  G.~J.~Olmo, D.~Rubiera-Garcia, and H.~Sanchis-Alepuz,
  Geonic black holes and remnants in Eddington-inspired Born-Infeld gravity,
  Eur.\ Phys.\ J.\ C {\bf 74}, 2804 (2014).
 
  \bibitem{BeltranJimenez:2017doy} 
  J.~Beltran Jimenez, L.~Heisenberg, G.~J.~Olmo, and D.~Rubiera-Garcia,
  Born-Infeld inspired modifications of gravity,
  Phys.\ Rept.\  {\bf 727}, 1 (2018).
  
\bibitem{Olmo:2016fuc} 
  G.~J.~Olmo, D.~Rubiera-Garcia, and A.~Sanchez-Puente,
  Impact of curvature divergences on physical observers in a wormhole space–time with horizons,
  Classical\ Quantum\ Gravity\  {\bf 33}, 115007 (2016).
  
\bibitem{Olmo:2015bya} 
  G.~J.~Olmo, D.~Rubiera-Garcia, and A.~Sanchez-Puente,
  Geodesic completeness in a wormhole spacetime with horizons,
  Phys.\ Rev.\ D {\bf 92}, 044047 (2015).

\bibitem{Olmo:2013mla} 
  G.~J.~Olmo and D.~Rubiera-Garcia,
  Semiclassical geons at particle accelerators,
  J. Cosmol. Astropart. Phys. 02 (2014) 010.

\bibitem{Olmo:2015dba} 
  G.~J.~Olmo, D.~Rubiera-Garcia and A.~Sanchez-Puente,
  Classical resolution of black hole singularities via wormholes,
  Eur.\ Phys.\ J.\ C {\bf 76}, 143 (2016).

\bibitem{Giveon:2004yg} 
  A.~Giveon, B.~Kol, A.~Ori, and A.~Sever,
  On the resolution of the timelike singularities in Reissner-Nordstr\"om and negative mass Schwarzschild,
  J. High Energy Phys. 08 (2004) 014.
  
\bibitem{Lobo:2013ufa} 
  D.~Rubiera-Garcia, G.~J.~Olmo, and F.~S.~N.~Lobo,
  Quadratic Palatini gravity and stable black hole remnants,
  Springer Proc.\ Phys.\  {\bf 170}, 283 (2016).

\bibitem{Lobo:2014fma} 
  F.~S.~N.~Lobo, G.~J.~Olmo, and D.~Rubiera-Garcia,
  Microscopic wormholes and the geometry of entanglement,
  Eur.\ Phys.\ J.\ C {\bf 74}, 2924 (2014).

\bibitem{Chen:2014jwq} 
  P.~Chen, Y.~C.~Ong, and D.~h.~Yeom,
  Black Hole Remnants and the Information Loss Paradox,
  Phys.\ Rept.\  {\bf 603}, 1 (2015).

\bibitem{Cardoso:2017cqb} 
  V.~Cardoso, and P.~Pani,
  Tests for the existence of black holes through gravitational wave echoes,
  Nature\ Astronomy\  {\bf 1}, 586 (2017).

\bibitem{Macedo:2018yoi} 
  C.~F.~B.~Macedo, T.~Stratton, S.~Dolan, and L.~Crispino, C.B.,
  Spectral lines of extreme compact objects,
 Phys.\ Rev.\ D {\bf 98}, 104034 (2018).

\bibitem{Olmo:2011uz} 
  G.~J.~Olmo,
  Palatini Approach to Modified Gravity: f(R) Theories and Beyond,
  Int.\ J.\ Mod.\ Phys.\ D {\bf 20}, 413 (2011).

\bibitem{BeltranJimenez:2019acz} 
  J.~Beltran Jimenez and A.~Delhom,
  Ghosts in metric-affine higher order curvature gravity,
  arXiv:1901.08988 [gr-qc].

\bibitem{Afonso:2017bxr} 
  V.~I.~Afonso, C.~Bejarano, J.~Beltran Jimenez, G.~J.~Olmo, and E.~Orazi,
  The trivial role of torsion in projective invariant theories of gravity with non-minimally coupled matter fields,
  Classical\ Quantum\ Gravity\  {\bf 34}, 235003 (2017).
  
\bibitem{Hawk} 
S. Hawking, 
Gravitationally collapsed objects of very low mass,
Mon. Not. R. Astron. Soc. {\bf 152}, 75 (1971).
 
\bibitem{Lobo:2013prg} 
  F.~S.~N.~Lobo, G.~J.~Olmo, and D.~Rubiera-Garcia,
  Semiclassical geons as solitonic black hole remnants,
  J. Cosmol. Astropart. Phys. 07 (2013) 011.

\bibitem{Wald} 
R. M. Wald,
\emph{General Relativity} 
(The University of Chicago Press, Chicago, 1984).

\bibitem{Casti2005}
J. Casti\~neiras, L. C. B. Crispino, R. Murta, and G. E. A.
Matsas, Semiclassical approach to black hole absorption of
electromagnetic radiation emitted by a rotating charge,
Phys. Rev. D {\bf 71}, 104013 (2005).

\bibitem{Macedo2014} 
C. F. B. Macedo and L. C. B. Crispino, Absorption of planar
massless scalar waves by Bardeen regular black holes, 
Phys. Rev. D {\bf 90}, 064001 (2014).

\bibitem{chm}
L. C. B. Crispino, A. Higuchi, and G. E. A. Matsas,
Low-frequency absorption cross section of the
electromagnetic waves for extreme Reissner-Nordstr\"om
black holes in higher dimensions, Phys. Rev. D {\bf 82},
124038 (2010).

\bibitem{och2011}
E. S. Oliveira, L. C. B. Crispino, and A. Higuchi, Equality
between gravitational and electromagnetic absorption cross
sections of extreme Reissner-Nordstrom black holes, Phys.
Rev. D {\bf 84}, 084048 (2011).

\bibitem{bodc}
C. L. Benone, E. S. de Oliveira, S. R. Dolan, and L. C. B. Crispino, 
Absorption of a massive scalar field by a charged
black hole, Phys. Rev. D {\bf 89}, 104053 (2014);
C. L. Benone, E. S. de Oliveira, S. R. Dolan, and L. C. B. Crispino, 
Addendum to absorption of a massive scalar field
by a charged black hole, Phys. Rev. D {\bf 95}, 044035 (2017).

\bibitem{ldc}
L. C. S. Leite, S. R. Dolan, and L. C. B. Crispino, Absorption
of electromagnetic and gravitational waves by Kerr
black holes, Phys. Lett. B {\bf 774}, 130 (2017);
L. C. S. Leite, S. Dolan, and L. C. B. Crispino, 
Absorption of electromagnetic plane waves by rotating black holes,
Phys. Rev. D {\bf 98}, 024046 (2018).

\bibitem{bc2016}
C. L. Benone and L. C. B. Crispino, Superradiance in static
black hole spacetimes, Phys. Rev. D {\bf 93}, 024028 (2016).

\bibitem{lbc2017}
L. C. S. Leite, C. L. Benone, and L. C. B. Crispino, Scalar
absorption by charged rotating black holes, Phys. Rev. D {\bf 96},
044043 (2017).

\bibitem{Futterman:1988ni} 
  J.~A.~H.~Futterman, F.~A.~Handler, and R.~A.~Matzner,
  \emph{Scattering From Black Holes}
   (Cambridge University Press, Cambrige, 1988).
	
\bibitem{Kokkotas:1999bd} 
  K.~D.~Kokkotas and B.~G.~Schmidt,
  Quasinormal modes of stars and black holes,
  Living Rev.\ Rel.\  {\bf 2}, 2 (1999).

\bibitem{Cardoso:2014sna} 
  V.~Cardoso, L.~C.~B.~Crispino, C.~F.~B.~Macedo, H.~Okawa, and P.~Pani,
  Light rings as observational evidence for event horizons: long-lived modes, ergoregions and nonlinear instabilities of ultracompact objects,
  Phys.\ Rev.\ D {\bf 90}, 044069 (2014).
	
\bibitem{Cardoso:2008bp} 
  V.~Cardoso, A.~S.~Miranda, E.~Berti, H.~Witek, and V.~T.~Zanchin,
  Geodesic stability, Lyapunov exponents and quasinormal modes,
  Phys.\ Rev.\ D {\bf 79}, 064016 (2009).

\bibitem{Gurvitz:1988zz} 
  S.~A.~Gurvitz,
  Novel approach to tunneling problems,
  Phys.\ Rev.\ A {\bf 38}, 1747 (1988).

\bibitem{Olmo:2017fbc} 
  G.~J.~Olmo, D.~Rubiera-Garcia, and A.~Sanchez-Puente,
  Accelerated observers and the notion of singular spacetime,
  Classical \ Quantum\ Gravity\  {\bf 35}, 055010 (2018).

\bibitem{Silvestri:2013ne} 
  A.~Silvestri, L.~Pogosian, and R.~V.~Buniy,
  Practical approach to cosmological perturbations in modified gravity,
  Phys.\ Rev.\ D {\bf 87}, 104015 (2013)

  
  \bibitem{Bazeia:2017gub} 
  D.~Bazeia, M.~A.~Marques, and R.~Menezes,
  Twinlike Models for Kinks, Vortices and Monopoles,
  Phys.\ Rev.\ D {\bf 96}, 025010 (2017)

  
\bibitem{Bazeia:2011hj} 
  D.~Bazeia and R.~Menezes,
  New results on twinlike models,
  Phys.\ Rev.\ D {\bf 84}, 125018 (2011)
  
  
  \bibitem{Lombriser:2015sxa}
  L.~Lombriser and A.~Taylor,
  Breaking a Dark Degeneracy with Gravitational Waves,
  J. Cosm. Astropart. Phys. 03 (2016) 031.

\bibitem{Lombriser:2016yzn}
  L.~Lombriser and N.~A.~Lima,
  Challenges to Self-Acceleration in Modified Gravity from Gravitational Waves and Large-Scale Structure,
  Phys.\ Lett.\ B {\bf 765}, 382 (2017).

 \bibitem{Baker:2017hug}
  T.~Baker, E.~Bellini, P.~G.~Ferreira, M.~Lagos, J.~Noller and I.~Sawicki,
  Strong constraints on cosmological gravity from GW170817 and GRB 170817A,
  Phys.\ Rev.\ Lett.\  {\bf 119},  251301 (2017).

\bibitem{Sakstein:2017xjx}
  J.~Sakstein and B.~Jain,
  Implications of the Neutron Star Merger GW170817 for Cosmological Scalar-Tensor Theories,
  Phys.\ Rev.\ Lett.\  {\bf 119},  251303 (2017).

  \bibitem{Creminelli:2017sry}
  P.~Creminelli and F.~Vernizzi,
  Dark Energy after GW170817 and GRB170817A,
  Phys.\ Rev.\ Lett.\  {\bf 119}, 251302 (2017).

   \bibitem{Ezquiaga:2017ekz}
  J.~M.~Ezquiaga and M.~Zumalac\'arregui,
  Dark Energy After GW170817: Dead Ends and the Road Ahead,
  Phys.\ Rev.\ Lett.\  {\bf 119}, 251304 (2017).

\bibitem{Ezquiaga:2018btd} 
  J.~M.~Ezquiaga and M.~Zumalac\'arregui,
  Dark Energy in light of Multi-Messenger Gravitational-Wave astronomy,
  Front.\ Astron.\ Space Sci.\  {\bf 5}, 44 (2018).

  \bibitem{Zerilli:1974ai} 
  F.~J.~Zerilli,
  Perturbation analysis for gravitational and electromagnetic radiation in a reissner-nordstroem geometry,
  Phys.\ Rev.\ D {\bf 9}, 860 (1974).

\bibitem{CHO:2009} 
L.~C.~B.~Crispino, A. Higuchi, and E.~S.~de Oliveira,  
Electromagnetic absorption cross section of Reissner-Nordstr\"om black holes revisited, 
Phys.\ Rev.\ D {\bf 80}, 104026 (2009).

\bibitem{CDHO:2014} L.~C.~B.~Crispino, S.~R.~Dolan, A.~Higuchi, and E.~S.~de Oliveira, 
Inferring black hole charge from backscattered electromagnetic radiation, 
Phys.\ Rev.\ D {\bf 90}, 064027 (2014).

\bibitem{CDHO:2015} L.~C.~B.~Crispino, S.~R.~Dolan, A.~Higuchi, and E.~S.~de Oliveira, 
Scattering from charged black holes and supergravity, 
Phys.\ Rev.\ D {\bf 92}, 084056 (2015).

\bibitem{Latorre:2017uve} 
  A.~Delhom, G.~J.~Olmo and M.~Ronco,
  Observable traces of non-metricity: new constraints on metric-affine gravity,
  Phys.\ Lett.\ B {\bf 780}, 294 (2018).

    
\end{thebibliography}

\end{document}